# Visible light induced ocular delayed bioluminescence as a possible origin of negative afterimage

2011
DOI: 10.1016/j.jphotobiol.2011.03.011

Bókkon I.*[1,2] Vimal R.L.P. [2,3,4,5] Wang C.[6] Dai J.[6] Salari V.[7,8] Grass F.[9] Antal I.[10]

[1]*Doctoral School of Pharmaceutical and Pharmacological Sciences, Semmelweis University, Budapest, Hungary;*

[2]*Vision Research Institute, 25 Rita Street, Lowell, MA 01854 USA and 428 Great Road, Suite 11, Acton, MA 01720 USA;*

[3]*Dristi Anusandhana Sansthana, A-60 Umed Park, Sola Road, Ahmedabad-61, Gujrat, India;*

[4]*Dristi Anusandhana Sansthana, c/o NiceTech Computer Education Institute, Pendra, Bilaspur, C.G. 495119, India;*

[5]*Dristi Anusandhana Sansthana, Sai Niwas, East of Hanuman Mandir, Betiahata, Gorakhpur, U.P. 273001, India;*

[6]*Wuhan Institute for Neuroscience and Neuroengineering, South-Central University for Nationalities, China;*

[7]*Kerman Neuroscience Research Center (KNRC), Kerman, Iran*

[8]*Afzal Research Institute, Kerman, Iran*

[9]*Department for Biological Psychiatry, Medical University of Vienna;*

[10]*Department of Pharmaceutics, Semmelweis University, Budapest, Hungary;*

*Corresponding author: István BÓKKON
Corresponding author's Email: bokkoni@yahoo.com
Corresponding author's Address: H-1238 Budapest, Lang E. 68. Hungary
Corresponding author's Phone: +36 20 570 6296
Corresponding author's Fax: + 36 1 217-0914

**Abstract**
The delayed luminescence of biological tissues is an ultraweak reemission of absorbed photons after exposure to external monochromatic or white light illumination. Recently, Wang, Bókkon, Dai and Antal (Brain Res. 2011) presented the first experimental proof of the existence of spontaneous ultraweak biophoton emission and visible light induced delayed ultraweak photon emission from in vitro freshly isolated rat's whole eye, lens, vitreous humor and retina. Here, we suggest that the photobiophysical source of negative afterimage can also occur within the eye by delayed bioluminescent photons. In other words, when we stare at a colored (or white) image for few seconds, external photons can induce excited electronic states within different parts of the eye that is followed by a delayed reemission of absorbed photons for several seconds. Finally, these reemitted photons can be absorbed by non-bleached photoreceptors that produce a negative afterimage. Although this suggests the photobiophysical source of negative afterimages is related retinal mechanisms, cortical neurons have also essential contribution in the interpretation and modulation of negative afterimages.

## 1. Introduction

When we stare at a colored (or white) image for several seconds and then look at a blank screen, we can see a complementary negative afterimage. This emerged negative afterimage is the same shape as the original image but different colors. Until now, a general assumption of negative afterimages is based on the photopigment-bleaching hypothesis [38,39,40]. Namely, after we stared at a colored image for few seconds, bleached photoreceptors are not sensitive to relevant visible photon stimuli compared to those photoreceptors that are not affected. Nevertheless, there are disagreements about photopigment bleaching hypothesis. Because we can see negative afterimages with closed eyes in a dark room, the photopigment bleaching idea does not make clear that where the source of negative afterimages is, i.e., what generates this long-lasting signals that makes fixed pictures inside our closed eyes without any external photonic source that is interpreted by neural mechanisms. Based on the Wang, Bókkon, Dai, and Antal [10] experiments (they proved spontaneous and visible light induced biophoton emission from *in vitro* freshly isolated rat's whole eye, lens, vitreous humor and retina) and light-induced photon reemission (delayed luminescence) phenomenon [12,13,14,15], here, we put forward a new photobiophysical interpretation of negative afterimage formation.

## 2. Rods and cones

As summarized in [16], "In the human retina, there are three types of photoreceptive pathways [17] (Dacey et al., 2005): (i) short-, medium- and long-wavelength-sensitive cone photoreceptors for day vision; (ii) rod photoreceptors for night vision; and (iii) melanopsin-expressing photosensitive ganglion cells for the functions of circadian photoentrainment and pupil constriction …". In the vertebrate retina, rods mediate night [18] vision and cones mediate daylight vision. A rod cell in the eye can perceive and transform a single photon (the smallest unit of energy) of light into a neural signal [19,20]. Still, in complete darkness, cones require the coincident absorption of some photons to generate a detectable signal [21,22,23]. The light-sensitivity of cones is $10^2$ times lower than that in rods and the photoresponse kinetics are much faster in cones [24]. The cones are also referred to as 'red', 'green', and 'blue' cones, but each cone class does not code the perception of a single color. Each cone pigment is an opsin, but the opsins have different amino acid sequences that restrict the accompanying chromophore so that it preferentially absorbs only one part of the visible spectrum.

The cones and rods are not uniformly distributed in the retina. The fovea is densely packed with cones. Around the fovea there are cones and rods, but there are fewer and fewer cones as the distance from the fovea increases. The region of the retina closest to the lens muscles contains primarily rods. In the central of the fovea (foveola) there are no rods at all.

Pokorny and his colleagues [25] determined the hue perceptions of paper color samples for a wide range of light levels, including very low light levels where rods alone mediated vision. They suggested that rods and L-cones might mediate color percepts at the intermediate light levels. In their experiments, at the three lowest light levels there were distinct color appearances mediated exclusively by rods.

"Human color vision is based on three light-sensitive pigments. …The red and green pigments show 96 percent mutual identity but only 43 percent identity with the blue pigment. Green pigment genes vary in number among color-normal individuals and, together with a single red pigment gene, are proposed to reside in a head-to-tail tandem array within the X chromosome" [26].

It had been thought that each mammalian cone photoreceptor could contain only a single type of photopigment. However, immunohistochemical evidence indicated that cones of several mammals coexpress both an S-pigment and an M/L-pigment in various animals such as mice, rabbits, hamsters, pigs, and also in humans [27,28,29].

## 3. Negative afterimages

Entoptic phenomena (Entoptic: from the Greek etymology: 'within vision') are visual effects whose sources are anywhere within the visual system [30]. The phosphenes and afterimages are known to be entoptic visual sensations. Historically, the retinal versus central origin of after-images has been debated [31,32]. There are two kinds of afterimages: negative afterimage, and positive afterimage [33].

Positive afterimages can occur because of the biochemical reactions are not instantaneous and the visual stimulus remains in the brain for a certain amount of time. A positive afterimage usually is a small fraction of a second. Positive afterimages appear in the same color as the original image. A positive afterimage, which probably reflects persisting activity of the visual system, is immediately followed by a negative afterimage that reflects the adaptation process and can last for much longer time.

The negative afterimage is the other kind of afterimages. If you stare at a colored image for some time and then look at a blank screen, you can see a complementary afterimage. A negative afterimage is the same shape as the original image but different colors. Colors, which are opposite to one another on the hue circle (Fig.1), are called complementary. If two complementary colors are mixed in the proper proportion, they generate a neutral color such as grey, white, or black.

Unfortunately, several textbooks say if the original image was red, the negative afterimage will be green, if the image was green, the negative afterimage will be red, and blue areas become yellow, and yellow areas become blue [34]. One could argue that this description may not be consistent with the subjective experiences reported by some other subjects. For example, cyan is a binary color that several subjects would use the phrase greenish-blue to describe. For some subjects, staring at red generates a cyan afterimage, staring at green generates a magenta afterimage, and staring at blue generates a yellow afterimage (and black and white also reverses) [33,35,36]. Chromatic adaptation results changes in hue, saturation and brightness [37] this may partly explain the subjective differences in negative afterimages.

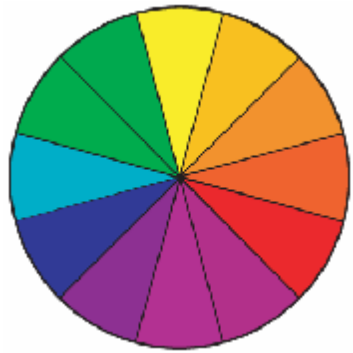

**Fig. 1**. The hue circle

### 4. Photopigment-bleaching, retinal and cortical mechanisms for negative afterimages

To date, a broadly accepted explanation of negative afterimages is based on the photopigment-bleaching hypothesis [38,39,40]. That is, bleached photoreceptors are not sensitive to relevant visible photon stimuli compared to those photoreceptors that are not affected.

However, there are contradictions about photopigment bleaching hypothesis. Several studies support that the origin of negative afterimages is not due to bleaching, rather because of other mechanisms [41,42,43,44,45]. For example, Craik [41] showed that positive and negative afterimages could be observed following adaptation to a 60 W illumination despite the fact that the eye was pressure-blinded during the 2 min adaptation. Craik concluded that this was proof of a photochemical basis of negative afterimages. In 1972 Loomis [43] also showed that bleaching is not the basis of certain afterimages. He demonstrated that after long duration illumination, afterimages were correlated with visual appearance, and were rather independent of bleaching. According to Lack' experiments [42], binocular rivalry does not reduce afterimage formation. In binocular rivalry, different images are presented to the two eyes and then the images are visible only alternately. In 2004, Hofstoetter, Koch and Kiper [46] demonstrated that motion-induced blindness does not significantly reduce the duration or the strength of negative afterimages. The motion-induced blindness can occur if a salient static target spontaneously fluctuates in and out of our visual awareness while surrounded by a random moving visual pattern [47]. The negative afterimage does not transfer between eyes [42], but [48] (Virsu and Laurinen, 1977) reported long-lasting afterimages caused by neural adaptation. All of these studies suggest that afterimages are also retinal phenomena.

Recently, Silvanto and his co-workers [49] performed a few exciting TMS (Transcranial magnetic stimulation) experiments. TMS induction was applied on the occipital cortex following 30 sec visual adaptation to a uniform color. TMS stimulation could produce phosphenes from the occipital cortex to take on the color qualities of the adapting color. For instance, if subjects adapted to a green color, they perceived a red negative afterimage into which the application of TMS induced green phosphenes. This phenomenon was found with red, green, blue and yellow adaptation colors. The negative afterimages lasted a mean duration of 69 sec. The size and shape of cortical phosphenes were not affected by visual adaptation. The perception of colored phosphenes stayed intense in every subject after the negative afterimage had paled. Since the perception of cortical phosphenes was not dependent on the presence of a visual negative afterimage, it supports that the negative afterimages have to have a retinal basis. However, one could argue that if phosphene-area does not overlap the negative afterimage's field of view (FOV), then this conclusion may be incorrect because they may be processed independently at different eccentricities.

Furthermore, several studies seem to imply an additional cortical role to the formation of negative afterimages [32,50,51,52]. For example, Tsuchiya and Koch [45], using the continuous flash suppression technique, reported: "the strength of the negative afterimage of an adaptor was reduced by half when it was perceptually suppressed by input from the other eye. […] Our results imply that formation of afterimages involves neuronal structures that access input from both eyes but that do not correspond directly to the neuronal correlates of perceptual awareness. […] One implication of our finding is that failure of interocular transfer and failure of reduction of afterimage intensity by partial suppression does not imply that structures that have access to information from both eyes, such as visual cortex, are not involved in the formation of negative afterimages."

We should also mention that phosphenes can be induced in blind people, for example, in the retina or V1 cortex, by electric or magnetic stimulations. However, complementary negative afterimages cannot be elicited in the retina or V1 cortex of blind people via electric or magnetic stimulations. This can also support the notion that the source of negative afterimages are placed somewhere within the eyes.

To sum up, negative afterimages most likely have a photophysical (photochemical) basis that is related to retinal mechanisms in addition to cortical neurons that participate in the modulation and interpretation of negative afterimages.

**5. Biophotons and delayed biophoton emission**

Ultraweak spontaneous photons (also called biophotons) are constantly emitted by all living systems without any external excitation [53,54,55,56,57,58,59,60]. The source of biophotons is due to the different biochemical reactions, primarily bioluminescent radical reactions of Reactive oxygen species (ROS) and Reactive nitrogen species (RNS) and the simple cessation of excited states. The key source of biophotons derives from oxidative metabolism of mitochondria and lipid peroxidation [6,8]. Neural cells also emit continuously biophotons during their ordinary metabolism [61,62]. In vivo intensity of biophoton emission from a rat's brain correlated with cerebral energy metabolism, EEG activity, cerebral blood flow, and oxidative stress [63].

According to Bókkon et al. [11], the real biophoton intensity inside cells can be significantly higher than the one expected from the measurements of ultraweak bioluminescence (usually measured some centimeters away from the cells). Since photons are strongly scattered and absorbed in cellular systems, the corresponding intensity of biophotons within the living systems and cells can even be two orders of magnitude higher.

Recently, Bókkon hypothesized [1] a new redox molecular mechanism for phosphene lights. That is, phosphene lights are because of the transient overproduction of free radicals and excited species that can produce an excess bioluminescent photon emission in diverse parts of the visual system. Above a distinct threshold this excess bioluminescent photon emission can emerge as phosphene lights, and the brain interprets these retinal bioluminescent photons so, as if they would originate from the external world.

In addition, Bókkon and Vimal [9] put forward a novel interpretation related to retinal discrete dark noises in rods. They pointed out that not only retinal phosphenes but also the discrete dark noise of rods can be due to the natural redox related (free radicals) bioluminescent biophotons in the retina. Since lipid peroxidation is a natural biochemical process in cells and also in retinal membrane [4,5], and natural lipid peroxidation is one of the key sources of ultraweak biophotons, and the photoreceptors have the highest polyunsaturated fatty acid concentration and oxygen consumption in the body, there can be a continuous, low level bioluminescent photon emission in the retina without external photonic stimulus [2,3,6,7,8].

However, the novel concept of retinal discrete dark noise [9] can be supported by experiments of Wang, Bókkon, Dai and Antal [10]. They presented the first experimental results of the existence of spontaneous and visible light induced (also called as delayed luminescence) ultraweak photon emission from *in vitro* freshly isolated rat's whole eye, lens, vitreous humor and retina [10].

Delayed luminescence (DL, also called as delayed biophoton emission or light-induced ultraweak photon emission) is the long-term ultraweak reemission of optical photons from various cells, organisms, and other material if they were illuminated by a white or monochromatic light [12,13,14]. The DL intensity is significantly lower than the well known fluorescence or phosphorescence. Although externally measured intensity of spontaneous biophoton emission from mammalian is extremely low intensity (from a few to up to some hundred photons/($cm^2$ x s), it can be well induced by illumination [14,15]. The decay time of delayed luminescence is dependent on the physiological conditions of the samples and kinds of the tissues as well as the conditions of illumination such as intensity, duration, and spectral distribution [14]. Delayed luminescence has hyperbolic behavior of the time decay [13].

Various papers demonstrated that the delayed luminescence is a sensitive indicator of the physiological state of cells that is closely connected to the differentiation stage of the biological system [14,64,65,66,67,68]. Lately, Grass and Kasper [69] demonstrated that DL is essentially dependent on the physiological conditions of the samples. They found out that whole blood does not reemit any photon after external UV illumination. In contrast, the same procedure with serum gives intense photon reemission for more than 10 minutes.

The light-induced photon emissions from the biological samples are not a random but can be a cooperative phenomenon. According to Simanonok's experiments [70], collagen fibers of tendon exhibit a fiber optic property of axially conducting light. Recently, Ho and his co-workers [12] studied the behavior of DL in a stable collagenous tissue (bovine Achilles' tendon), which has a hierarchical, fractal fibrous structure. The DL characteristics largely depended upon structural organization. They proposed [12] that DL comes from the excitation and subsequent decay of collective electronic states whose properties depend on the organized structure of the system. Namely, extremely long decay times of DL are due to the interrelated molecular levels that are connected to an ordered spatial structure.

## 6. Ocular delayed luminescence as a possible photbiophysical source of negative afterimage

It is well known if we stare at a colored image for a time and then look at a blank white screen, we can see a complementary negative afterimage. This negative afterimage is the same shape as the original image but different in color. For instance, if the original image was red, the negative afterimage will be cyan (or green) depending on stimulus and adaptation condition [33,36]; see also [37] for the change of hue, saturation, and brightness after viewing lights steadily.

We can see long-lasting negative afterimages in both eyes-open and eyes-closed situations depending on stimulus and adaptation conditions. Under eye-open condition, you look at white screen with your eyes open after staring for example the red square for long time and you experience cyan (or green) negative after image. One might explain this eyes-open negative after images by usual photopigment-bleaching and adaptation-causing-fatigue hypothesis. According to the photopigment-bleaching hypothesis, some of the long wavelength sensitive (LWS) photoreceptors in the eye are bleached (chromophore retinal changes conformation from 11-cis to all-trans) by red-square reflecting (or emitting) long wavelength light. After staring long time, when you view white screen (that contains all wavelengths), middle and short wavelength lights of white screen are absorbed by unbleached and unadapted photoreceptors, which then lead you experience cyan (or green) negative after image.

Nevertheless, if your eyes are closed after staring the red square for long time and you experience cyan (or green) negative after image, then how to explain it? Since we can see negative afterimages not only with open eyes but also closed eyes without any external photonic sources in a dark room, the photopigment bleaching hypothesis cannot explain that. In other words, where is the source of long-lasting negative afterimages with closed eyes without any external photonic sources? That is, what produces this long-lasting signal that makes static pictures within our eyes, which is interpreted by higher neural mechanism in the brain?

However, Wang et al.'s [10] experiments may suggest that negative afterimages can be due to the photons of delayed luminescence within the eyes activating non-bleached photopigments. During normal vision, the eyes are continuously exposed to ambient powerful photons that pass through various parts of the eyes. When a subject stares at an image for several seconds, a small fraction of external photons can be absorbed in various parts of eyes, and a little later reemitted within the eyes. This phenomenon is called as delayed (bio)luminescence [13]. Next, reemitted delayed photons (that could produce an afterimage, see our Figure 2.) are absorbed by retinal non-bleached photopigments and then retinotopic electrical signals are conveyed to the V1, the V2, etc., and to higher levels of cognitive interpretations (similarly like external photons during vision). Thus the brain can interpret an afterimage (produced via delayed photons) as if they originate from the outside visual world. It is important to consider that under natural circumstances, ambient photons can induce much stronger delayed photon emission within the eyes than in the isolated and dark-adapted experiments [10].

Since, for example, external long wavelength photons bleached LWS photopigments, the released delayed long wavelength photons can be absorbed by middle wavelength sensitive (MWS) and short wavelength sensitive (SWS) blue and green photopigments that produce a weak green or cyan negative afterimage.

Recent experiments suggested that rods and L-cones might mediate color percepts at the intermediate light levels [25]. We need just two types of receptors to experience color. Since the light induced signals of rod photoreceptors in mammalian retina can be transmitted by chemical synapses to bipolar cells and via gap junctions to cone photoreceptors [71,72], it suggests that rods may also take a role in the formation of afterimages.

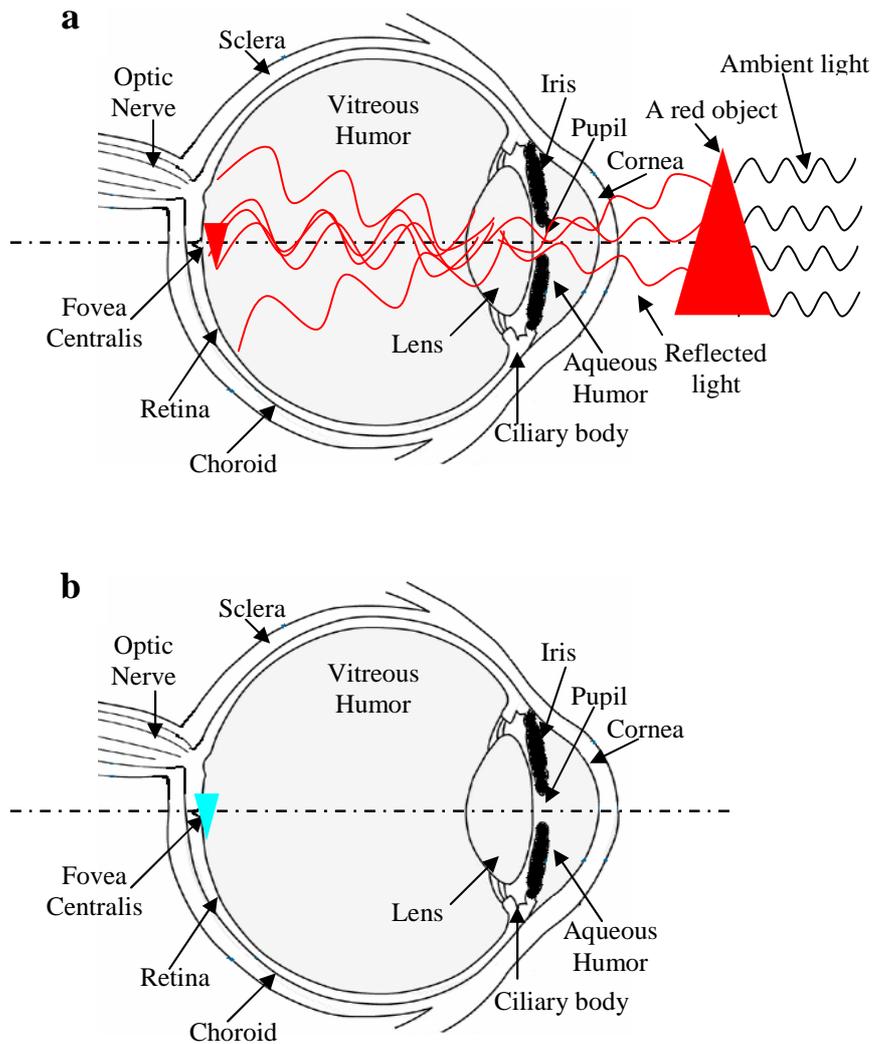

**Fig. 2. Simple figure of our hypothesis about photobiophysical source of negative afterimage.** (**2a**) During normal vision, when we stare at an image (red triangle) for several seconds, a small fraction of external photons can be absorbed in various parts of eyes. (**2b**) Following we stared at a colored image (red triangle), a tiny fraction of absorbed external photons, which were stored within the eyes, can be released (delayed luminescence) for several seconds and absorbed by non-bleached photoreceptors that produce an afterimage in the complementary color (tiny cyan triangle). Although several parts of the eye performed delayed photon emission [10], the retina can be the most possible candidates to form an afterimage by delayed luminescence in the eyes. However, methodologies need to be established to determine what the major candidate part can be accountable for creation of negative afterimages by delayed photons.

## 7. Black and white complementary afterimages

*Black and white objects in everyday life*

The spectrum of particle and electromagnetic radiations ranges from the extremely short wavelengths of cosmic rays and electrons to radio waves hundreds of kilometers in length. The color of light of a single wavelength is known as hue or a pure spectral color. Such pure colors are fully saturated and are seldom encountered outside the laboratory. The human eye does not function like a device for spectral analysis, i.e., the same color perception can be produced by different electromagnetic physical stimuli. For instance, a mixture of green and red light of the proper intensities appears exactly the same as spectral yellow, although it does not contain electromagnetic visible wavelengths corresponding to yellow.

Black or white, it's not an all or nothing case in everyday life. White objects absorb a definite fraction of incident visible photons that is dependent on the actual pigment and the shininess of the surface [73]. Black objects also reflect a definite fraction of incident electromagnetic photons for the same reasons.

Perceived white light is a mixture of all colors (namely, mixture of all visible electromagnetic photons). White things are white because the most of the light that falls on them is reflected by the material. Black objects absorb light of all frequencies but a little light is reflected from them. Since black-colored things, in everyday life, reflect a little light, this reflected light contains all visible frequencies because black objects absorb light of all frequencies [73].

*Black and white complementary afterimages*

The computer graphics could help us to understand the white and black complementary afterimages. Computer graphics uses combinations of just three lights (similarly to our eyes' cones) to produce the colors on a screen using red, green, and blue phosphors (see our Figure 3). The interesting thing about this system is, white and black have the same hue and saturation, and the lightness is all that is different.

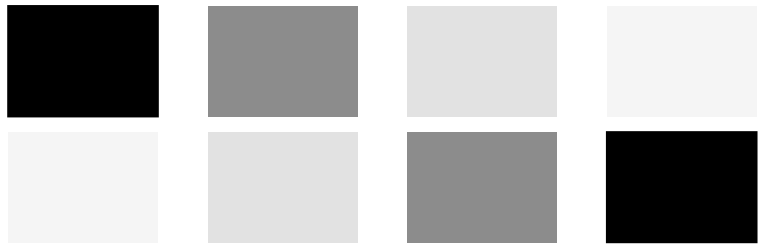

**Fig. 3.** White and black have the same hue and saturation, and the lightness is all that is different. The complementary afterimages of upper rectangles can be seen the under rectangles (or vice versa).

Now, let's see first the formation of black negative afterimages after white adaptation. Since white is a mixture of all colors, when we are staring at a white object for several seconds it can bleach the numbers of red, green, and blue photopigments. It is known that staring at red generates a cyan or green afterimage, green generates a magenta or red afterimage, and blue generates a yellow afterimage. So, when we are staring at a white object it generates simultaneously three kinds of afterimages (i.e. cyan or green, magenta or red, and yellow afterimages). However, mixture of cyan, magenta, and yellow form a black color (see central black area of Fig. 4b) because black is complementary to white (see central white area of Fig. 4a, which is the additive mixture of red, green, and blue). So, we are staring at a white object the transiently absorbed and stored external photons within the eyes can be released that can produce a black afterimage by a mixture of cyan, magenta, and yellow afterimages with the understanding that opposite of white is black.

When we are staring at a black object a little light is reflected to our eyes, but this reflection is also a mixture of all colors. But reflected photons from a black image hardly bleach red, green, and blue photopigments. However, during 30-60 sec adaptation, a considerably amount of external photons can be absorbed within the eyes. So, delayed photons can be absorbed by red, green, and blue photopigments that produce a white (mixed) afterimage.

It means that the main difference between black afterimage and white afterimage is that black negative afterimage (after white adaptation) produced by a mixture of cyan, magenta, and yellow afterimages (due to the significant bleaching) (Fig.4b), but white negative afterimage (after black adaptation) produced by a mixture of normal red, green, and blue photopigments (Fig.4a) (without significant bleaching).

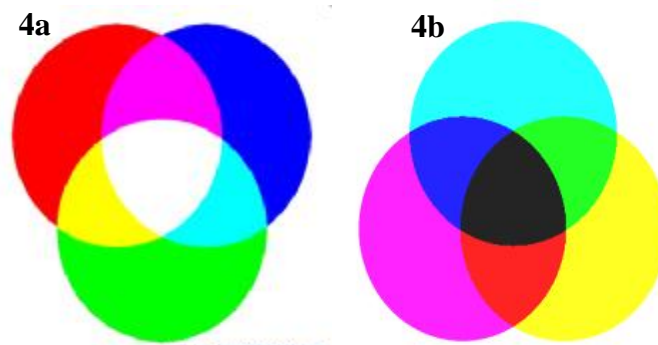

**Fig. 4.** **(a)** Additive color combinations. Primary colors: red, green, and blue. Additive color mixtures: red + green = yellow, green + blue = cyan, red + blue = megenta, and red + green + blue = white. Complementary colors: cyan, magenta, and yellow. **(b)** Subtractive color combinations. Primary colors: cyan, magenta and yellow. Complementary colors: red, green, and blue.

## 8. Implications on negative afterimage: wavelength shift, intensity and time delay

As it is known, when a molecule absorbs a photon it achieves an excited state. When the excited molecule relaxes to the ground state it can release a photon that generally occurs at longer wavelength (lower energy) than the absorbed wavelength of photon. During delayed luminescence (DL), following visible light illumination of cells, the photon emission spectrum in DL moves towards longer wavelengths [64]. These suggest that ocular reemitted delayed photons [10] can be essentially absorbed by middle wavelength sensitive (MWS) and long wavelength sensitive (LWS) photoreceptors. Besides, immunohistochemical evidence indicated that cones of several mammals coexpress both an S-pigment and an M/L-pigment in various animals such as mice, rabbits, hamsters, pigs, and humans [27,28,29]. Thus, the emergence of the afterimage may be due to the absorption of delayed biophotons by middle and long wavelength sensitive photoreceptors.

Because the intensity of delayed biophotons is dependent on several factors such as the actual physiological conditions of the samples and kinds of the tissues as well as the conditions of illumination such as intensity, duration, and spectral distribution, we can only speculate regarding the intensity of reemitted biophotons within different parts of the eye [10]. As mentioned above, a rod cell can perceive and transform a single photon of light into a neural signal [19,20], and in complete darkness, cones require the coincident absorption of some photons to produce a detectable signal [21,22,23]. Biophoton intensity is usually some hundred photons/(cm$^2$ x s) [14,15], but the intensity of delayed luminescence is usually 2-3 order larger than the spontaneous biophoton emission. We should consider that under natural circumstances, ambient photons can induce much stronger delayed photon emission within the eyes than in the isolated and dark-adapted experiments [10]. We should also consider that the DL characteristics largely depended upon structural organization of the system [12]. Although Wang et al. [10] demonstrated that various parts of *in vitro*, freshly isolated rat eye, including the lens, vitreous humor and retina as well as the whole eye, can display delayed biophoton emission, it is hardly possible that delayed biophotons of lens or vitreous humor could be a source of the afterimage, because induced visible photons can be absorbed easily. Thus, it is feasible that delayed biophotons of organized structural system of retina could produce the afterimage. Taken all together, we may calculate, at least in principle, with $10^4$ biophoton/sec in the retinal system during DL, which could be sufficient to produce an afterimage via retinal MWS and LWS cone photoreceptors.

Usually, an afterimage can remain for 30 seconds or longer. The delayed luminescence photonic signal can be observable for duration ranging from tens of milliseconds to several minutes in different systems [82]. Thus, the duration of an afterimage and the duration delayed luminescence within the eyes can be similar.

Finally, the circulating current in cones can recover within 3 sec following 50% of photopigment was bleached, while 15 min is required for a similar extent of recovery in human rods [80,81]. It means that human cones remain responsive under intensely light conditions but rods are not. In addition, afterimages usually became visible about 1-2 sec after the initial display. However, these raise the possibility that the negative afterimage phenomenon is not dependent on the photopigment bleaching, as Loomis [43] showed it in 1972. During Loomis' experiments, after long term illumination, negative afterimages were correlated with visual appearance, and were quite independent of bleaching.

## 9. Summary

A broadly accepted explanation of negative (complementary) afterimages is based on the photopigment-bleaching hypothesis. However, there are several contradictions about this idea.

It is known that long-lasting negative afterimages can also be seen with closed eyes without any external photonic sources. Since we can see negative afterimages with closed eyes in a dark room, the photopigment bleaching hypothesis can not explain that where the source of negative afterimages is, i.e., what produces this long-lasting signals that makes static pictures inside our closed eyes that is interpreted by higher neural mechanism.

Based on the mentioned several papers and Wang, Bókkon, Dai, and Antal [10] experiments, here, we suggested a novel photobiophysical interpretation of negative afterimage formation. Namely, the photobiophysical source of negative afterimage may be originated from the delayed bioluminescence within the eyes. Delayed luminescence is the long-term weak re-emission of visible photons from cells (or other material) on being illuminated with a white or monochromatic light. Delayed luminescence within the eyes is nothing else as natural low intensity photon reemission in various parts of the eyes during vision.

According to our novel concept, when you stare at a colored (or white) image for some seconds external photons can be absorbed and induce excited electronic states within the eyes. Then, there can be a long-term delayed reemission of absorbed photons. Finally, these reemitted photons can be absorbed by non-bleached cones (or/and rods) that produce a complementary afterimage.

We have to emphasize again, although this suggested photobiophysical source of negative afterimages is related to the retinal mechanisms, cortical neurons can have also fundamental contribution in the modulation and interpretation of negative afterimages. For example, without early retinotopic cortical processes we could not see any negative afterimage, because early visual areas are essential for visual apperception [74].

Although currently we do not know the importance and possible roles of ambient light induced bioluminescent delayed photons (within different parts of the eyes) during vision, it should be considered in vision research in the future. Besides, methodologies need to be established to determine what the major candidate part is within the eyes, which can be accountable for creation of negative afterimages by delayed photons. It also should be investigated if emission spectrum of delayed photons could be similar to the excitation one in the eyes. Finally, further studies should be conducted, which not only confirm Wang et al.'s [10] experimental results, but extend this novel area of ultraweak biophotonic research in vision investigations.

***Footnote: Phototransduction cascade and visual cycle in vertebrate photoreceptors**

The visual photopigment (a member of the G protein-coupled receptor family) contains two distinct components, a colorless protein called opsin and a chromophore (chromophore covalently bound to opsin) derived from vitamin A, which is known as retinal [75]. The pigment protein in rods is called rhodopsin, while the pigment protein in cones is called iodopsin. Upon photon absorption, 11-*cis*-retinal (that binds to opsin) undergoes an isomerization to the all-*trans* form (that does not bind to opsi*n*), causing a conformational change in the photopigment [76]. When the chromophore undergoes rapid photoisomerization, the visual pigment is said to be "bleached," i.e., it is no longer able to absorb visible photons. This rapid photochemical isomerization triggers an activation of the G-protein (transducin) signal transduction cascade that transmits a visual signal from the photoreceptor to intermediate (bipolar and amacrine) and ganglion cells in the retina and thence to the brain [77,78]. The regeneration of the visual pigment occurs by a sequence of biochemical reactions referred to as the 'visual cycle'. The first step of this visual cycle takes place in the photoreceptor outer segment. All-*trans* retinal is reduced to all-*trans* retinol by the photoreceptor retinol dehydrogenase [79]. Next, retinol is removed from the outer segment and transported to the retinal pigment epithelium (RPE). Following enzymatic conversion of all-*trans*-retinoid to 11-*cis*-retinal in the bordering RPE, it is transported back to photoreceptor outer segments to regenerate rhodopsin.

**Conflict of interest**
The authors report no conflicts of interest. The authors alone are responsible for the content.

**Acknowledgments**
The authors (i) Bókkon I. gratefully acknowledges support of this work by BioLabor (www.biolabor.org), Hungary. His URL: http://bokkon-brain-imagery.5mp.eu; (ii) Vimal R.L.P. would like to thank VP-Research Foundation Trust and Vision Research Institute research Fund for the support. Vimal's URL: http://sites.google.com/site/rlpvimal/Home